\documentclass{INTERSPEECH2023}

\usepackage{fvextra}
\usepackage{graphicx}
\usepackage{array}
\usepackage{multirow}
\usepackage{booktabs}
\usepackage{algorithmic}
\usepackage{amsmath,amssymb,amsfonts}
\usepackage{booktabs}

 \interspeechcameraready

\title{Multi-pass Training and Cross-information Fusion for Low-resource
End-to-end Accented Speech Recognition}
\name{Xuefei Wang$^1$, Yanhua Long$^1$$^*$\thanks{Yanhua Long is the corresponding author, she is also with the Shanghai Engineering Research Center of Intelligent Education and Bigdata, Shanghai Normal University. The work is supported by the National Natural Science Foundation of China (Grant No.62071302).}, Yijie Li$^2$, Haoran Wei$^{3}$}

\address{
  $^1$Key Innovation Group of Digital Humanities Resource and Research, \\
  Shanghai Normal University, Shanghai, China\\
  $^2$Unisound AI Technology Co., Ltd., Beijing, China\\
  $^3$Department of ECE, University of Texas at Dallas, Richardson, TX 75080, USA}
\email{xuefei\_wang@163.com, yanhua@shnu.edu.cn, liyijie@unisound.com, haoran.wei@utdallas.edu}

\begin{document}

\maketitle
 
\begin{abstract}

Low-resource accented speech recognition is one of the important challenges faced by current ASR technology in practical applications. In this study, we propose a Conformer-based architecture, called Aformer, to leverage both the acoustic information from large non-accented and limited accented training data. Specifically, a general encoder and an accent encoder are designed in the Aformer to extract complementary acoustic information. Moreover, we propose to train the Aformer in a multi-pass manner, and investigate three cross-information fusion methods to effectively combine the information from both general and accent encoders. All experiments are conducted on both the accented English and Mandarin ASR tasks. Results show that our proposed methods outperform the strong Conformer baseline by relative 10.2\% to 24.5\% word/character error rate reduction on six in-domain and out-of-domain accented test sets.

\end{abstract}

\section{Introduction}
\label{sec:Intro}

In recent years, the performance of automatic speech recognition (ASR)
on high-resource languages has benefited enormously from neural models~\cite{Chan2016listen, weninger2022conformer, Chiu2018State}.
The excellent performance makes these ASR systems widely being used in variety of
commercial speech recognition products~\cite{baevski2020wav2vec, zhang2020pushing, hsu2021hubert}.
However, it is well-known that ASR system performance degrades significantly when
encountering accent speech, especially when these accents are not existing in the ASR
training dataset~\cite{Huang2004accent}.
Accent is a special way of pronunciation, which is influenced by the region,
the speaking style and the level of speaker's education~\cite{Bahdanau2016end}, etc.
Such as in remote regions or villages of south China, those accents are very different
and seriously affect the pronunciation of Mandarin. All these accent
variations lead to extremely expensive and time-consuming for
transcribing heavy accented speech. Therefore, in the literature,
there is a serious data sparsity problem of non-mainstream accented speech,
building high performance ASR system for low-resource accented speech is very
important and fundamental.

In the past few years, many research works have been explored for
improving accented speech recognition. Such as in~\cite{Wang2003Comparison, Arslan1999a,Mayfield2022Adaptation},
different model adaptation methods were proposed to handle the
non-native speech recognition. In ~\cite{Huang2014Multi},
they proposed a multi-accent deep acoustic model with an accent-specific
top layer and shared bottom hidden layers. While in~\cite{Shor2019Personalizing},
authors explored to finetune a particular subset of neural network layers
with limited accent data. Fine-tuning a pre-trained model using limited accent speech is a straightforward way to handle the accent issue~\cite{Tan2021AIS, Jicheng2021E2E, deng2021improving}.
In most recent years, many works focus on
extracting representative accent embedding to improve the accented ASR,
such as in \cite{gong2022layer}, they extracted the accent embedding
from a well-trained accent classifier to perform the layer-wise adaptation
of end-to-end ASR model; In \cite{li2018multi}, they just used one-hot embedding
to build a multi-dialect ASR system; And in \cite{li2021end}, they
designed a TTS auxiliary model to convert accent information
into a global style embedding for improving the
accent robustness of E2E ASR model. All these previous works
have been greatly boosted the performance of accented speech recognition,
however, most works still requires balanced or large amount of accented
speech training data. The works about how to leverage the available
large amount of non-accent training data to improve the low-resource
accented ASR system are still very limited.

In this paper, we aim to investigate using large amount of non-accented
training data to boost the performance of low-resource accented speech
recognition. Three contributions are explored: 1) a unified architecture
with both general and accent encoders are proposed. These encoders are
designed to integrate the general acoustic information that learnt from
large available non-accented training data and the accent-dependent acoustic
information that extracted from the limited low-resource accented speech;
2) a multi-pass training is explored to build relatively stable
accent and general encoders; 3) Three cross-information fusion methods are
exploited to effectively combine the information from both general and
accent encoders. All methods are improvements of the state-of-the-art
Conformer~\cite{gulati2020conformer} system. Experiments are performed on both accented English
and Mandarin ASR tasks. Results show that, compared with the Conformer baseline,
on the in-domain Indian and Guangdong accent test sets, our proposed methods
can achieve 10.6\% relative word error rate (WER) reduction and 15.6\% relative character error rate (CER) reduction, respectively.
On the out-of-domain accented test sets, we also obtain 10.2\% to 24.5\%
performance improvements on both the English and Mandarin accented ASR tasks.

\begin{figure*}[th]
  \centering
  \includegraphics[width=16cm]{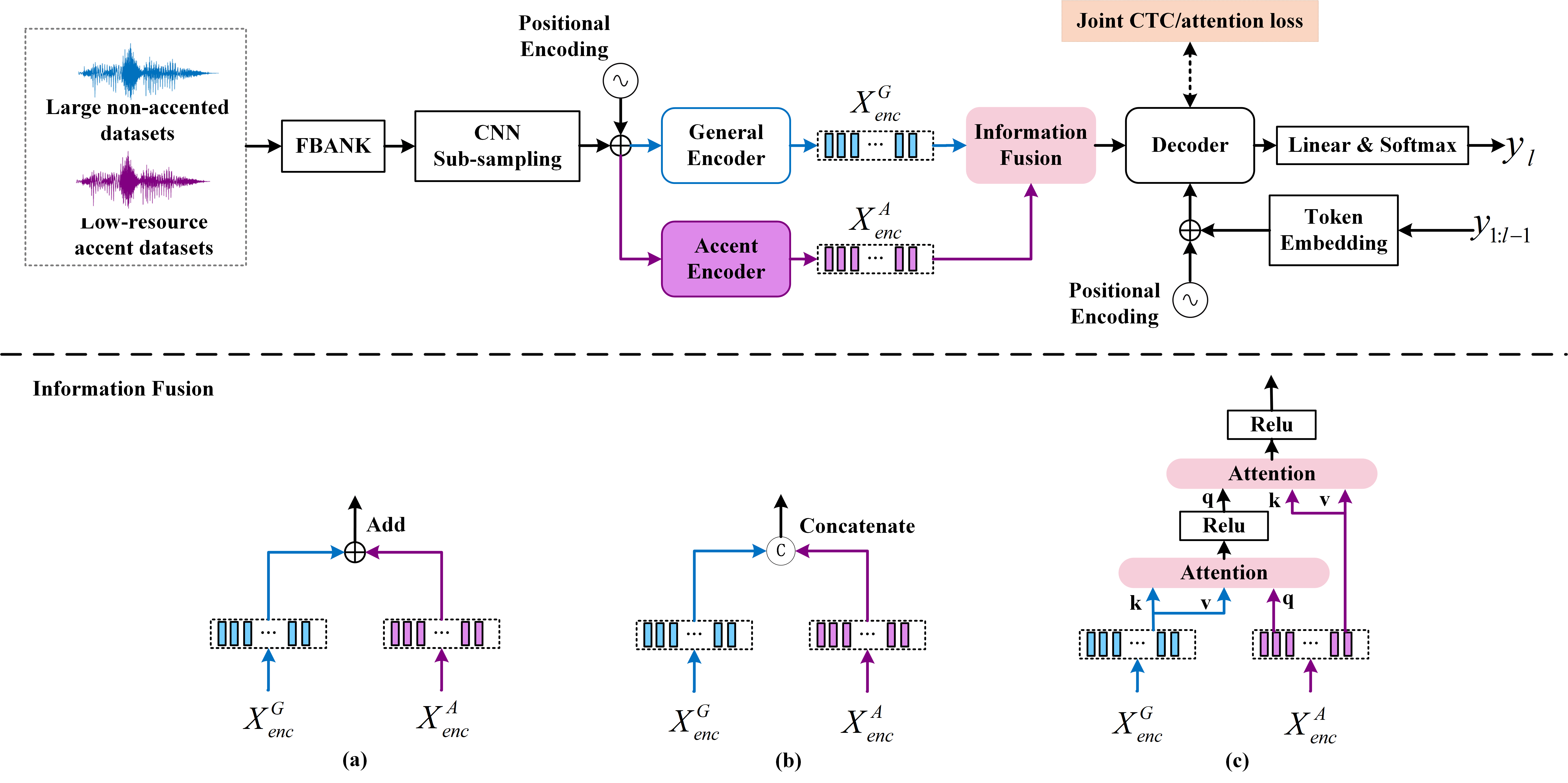}\\
  \caption{System architecture of the proposed Aformer framework.}
  \label{fig:Aformer}
\end{figure*}

\section{Conformer-based ASR}
\label{sec:Conformer}

In this paper, all our contributions are based on the convolution-augmented Transformer
(Conformer) E2E ASR model that has been recently proposed in ~\cite{gulati2020conformer}.
In order to improve the ability to capture locality of a sequence, Conformer inserts a
convolution layer into the transformer block~\cite{Guo2020RecentDO}. Because of its consistent excellent performance
over a wide range of ASR tasks~\cite{Zeineldeen2022conformer}, Conformer has been taken as the state-of-the-art
E2E ASR technique, more and more Conformer variants have been explored in recent years~\cite{deng2022confidence}.
The basic Conformer block consists mainly of four modules: the first feed-forward module (FFN1),
the multi-head self-attention module (MHSA), the convolution module (Conv) and another
feed-forward module (FFN2). Given an input sequence $x$, the output $y$ of one
Conformer block can be mathematically defined as:

\begin{equation}
\label{eq:conf}
\begin{aligned}
x_{\mathrm{FFN_{1}}} & =x +\frac{1}{2}\mathrm{FFN}(x),\\
x_{\mathrm{MHSA}} & = x_{\mathrm{FFN_{1}}} + \mathrm{MHSA}(x_{\mathrm{FFN_{1}}}),\\
x_{\mathrm{Conv}} & = x_{\mathrm{MHSA} } + \mathrm{Conv}(x_{\mathrm{MHSA}}),\\
x_{\mathrm{FFN_{2}}} & = x_{\mathrm{Conv}}+\frac{1}{2}\mathrm{FFN}(x_{\mathrm{Conv}}) ,\\
y &= \mathrm{Layernorm}(x_\mathrm{FFN_{2}})\\
\end{aligned}
\end{equation}

More details of the Conformer E2E ASR model can be referred to ~\cite{gulati2020conformer}.
During Conformer training, the following multi-task criteria
using interpolation of the CTC and attention cost is adopted~\cite{Xie2019BLHUC},
\begin{equation}
\label{loss}
  \mathcal{L} = (1-\lambda)\mathcal{L}_{att}+\lambda \mathcal{L}_{ctc}
\end{equation}
Where the task weight $\lambda$ is empirically set to 0.3 and
fixed throughout the experiments of this paper.

\section{Proposed Methods}
\label{sec:proposed}

In this section, we introduce the details of our proposed Aformer,
which is specially designed for improving the performance of
low-resource end-to-end accented speech recognition.
The whole model architecture is presented in Section~\ref{sec:Architecture},
followed by the description of multi-pass training in Section~\ref{sec:Multi-pass Training},
and the cross-information fusion methods are described in Section~\ref{sec:fusion}.

\subsection{Architecture}
\label{sec:Architecture}

The whole architecture of our proposed Aformer is illustrated in Fig.\ref{fig:Aformer}.
Compared with the standard Conformer in Section~\ref{sec:Conformer}, the only difference is the
purple highlighted two blocks: the accent encoder and the information fusion.
All other blocks are exactly the same as Conformer in Section~\ref{sec:Conformer}, including the
FBANK extraction, CNN sub-sampling, positional encoding, the general encoder,
and the decoder, etc.

Assuming the outputs of two encoders are $X_{enc}^{G}$ and $X_{enc}^{A}$
for the general and accent encoder, respectively. The information fusion block is
designed to effectively combine the different acoustic representations as
\begin{equation}
\label{fusion}
 X_{enc}^{F} = \mathrm{Fusion}(X_{enc}^{G},X_{enc}^{A})
\end{equation}
where the fusion methods are demonstrated in subfigure (a), (b) and (c) of Fig.\ref{fig:Aformer},
and they will be presented in detail in Section~\ref{sec:fusion}.
Finally, together with the token embedding,
the combined acoustic embedding $X_{enc}^{F}$ is then fed into the
decoder module to get the decoding outputs.

The principle behind the design of Aformer is that, we aim to leverage
the acoustic information in large amount of open-source non-accent training
speech to boost the low-resource accented speech recognition.
Therefore, in Aformer, we keep using the original Conformer encoder in
\cite{gulati2020conformer} as the general encoder to extract the general
accent-invariant acoustic context embedding,
while adding a much simpler network as the additional
accent encoder to learn the accent-dependent acoustic attributes from
extremely low-resource accented data. After a proper embedding fusion,
we expect these two different acoustic representatives will be well
integrated to improve the final accented E2E ASR system.

\subsection{Multi-pass Training}
\label{sec:Multi-pass Training}

The proposal of multi-pass training of Aformer is motivated by the heavy data
imbalance between non-accented and accented ASR training
speech. Under the low-resource accented ASR condition, the labeled
accented speech is normally a few minutes to tens of hours, these limited
data is not enough to well train a separate accent encoder. Therefore,
the multi-pass training is proposed, it aims to not only provide an
good initialization of the general and accent encoders, but also to
enable the Aformer concentrate on well training the specific encoder
at different training pass. The detail of Aformer multi-pass training
can be divided into three passes as follows:
\begin{itemize}
    \item \verb"Pre-training": Use the available large amount of non-accented
     training speech to train Aformer as the initialization model (A1).
    \item \verb"Accent encoder adaptation": Freeze the parameters of the
    general encoder, use the provided low-resource accent training data to
    only adapt the accent encoder of A1 to learn the specific accent
    acoustic characteristic. This adapted model is termed as A2.
    \item \verb"Re-training": Pooling both the non-accented and
     accent training data together to re-train the Aformer (A3) based on
     the adapted A2.
\end{itemize}
By performing the above three-pass model training, both the information in
the limited accented and large amount of non-accented training data are effectively
exploited, which helps the final Aformer A3
has the ability to well capture both the general acoustic context and accent-dependent
acoustic information for improving the low-resource accented E2E ASR system performance.

\subsection{Cross-information Fusion}
\label{sec:fusion}

In the proposed Aformer, how to combine the outputs of
its two different encoders is very important. In this study,
we investigate three ways to perform the information fusion
for validating the complementarity between two encoders'
outputs. Details are shown in Fig.\ref{fig:Aformer}
(a) to (c). Specifically, (a) and (b) are two traditional
information fusion methods that defined in Eq.(\ref{eq:a}) and (\ref{eq:b}):
the linear addition and the concatenation.

\begin{equation}
\label{eq:a}
 X_{enc}^{F}=\mathrm{Add}(X_{enc}^{G},X_{enc}^{A})
\end{equation}

\begin{equation}
\label{eq:b}
 X_{enc}^{F}=\mathrm{Concat}(X_{enc}^{G},X_{enc}^{A})
\end{equation}

Different from (a) and (b), in Fig.\ref{fig:Aformer}(c),
we propose a two-layer cross-attention structure to combine the
embeddings of the general and accent encoders as,
\begin{equation}
\label{eq:c}
\begin{aligned}
&X_{enc}^{M}=\mathrm{Relu}(\mathrm{Softmax}(\frac{Q_{enc}^{A} (K_{enc}^{G})^{T} }{\sqrt{d_{att}}})V_{enc}^{G}),\\
&X_{enc}^{F}=\mathrm{Relu}(\mathrm{Softmax}(\frac{Q_{enc}^{M} (K_{enc}^{A})^{T} }{\sqrt{d_{att}}})V_{enc}^{A}),\\
&Q_{enc}^{i}=W_{i}^{Q}X_{enc}^{i},K_{enc}^{i}=W_{i}^{K}X_{enc}^{i},V_{enc}^{i}=W_{i}^{V}X_{enc}^{i}
\end{aligned}
\end{equation}
Where $X_{enc}^{i}\in \left \{ X_{enc}^{G},X_{enc}^{A},X_{enc}^{M} \right \}$, $d_{att}$ is the
attention dimension. $Q_{enc}^{i}$, $K_{enc}^{i}$ and $V_{enc}^{i}$ are the query,
key and value linear projections on the encoder output $X_{enc}^{G}$, $X_{enc}^{A}$
and the output of the first layer cross-attention $X_{enc}^{M}$, respectively.
The projections are parameter matrices $W_{i}^{Q}$, $W_{i}^{K}$ and $W_{i}^{V}$.
This cross-attention structure is used to highlight
the complementary information between the general and accent-dependent
acoustic embeddings, making the combined encoder output $X_{enc}^{F}$ more
robust and representative.

\section{Experiments}
\label{sec:exp}

\subsection{Datasets}
\label{sec:data}

In this study, two language datasets are used to examine the effectiveness
of our proposed methods, one is English and the other is Mandarin.
To simulate low-resource accented ASR tasks, we use the
``train-clean-360"~\cite{Panayotov2015libri} as the non-accented English training
data, and randomly select 20 hours (hrs) English data with Indian (IN) accent
from the publicly available Common Voice~\cite{Ardila2020CommonVA} as
the limited accented training data. For Mandarin task,
the open-source Aishell~\cite{Bu2017AISHELL1AO} is taken as the non-accented
training data, while the accented Mandarin datasets are all collected from
a live speech service system of
Unisound corporation in China (https://www.unisound.com/).
20 hrs Mandarin with Guangdong (GD) accent is  selected as the
training data. Six accent test sets are used for system evaluation, including
the IN and GD in-domain test sets, and four out-of-domain test sets with
England (EN), Canada (CA), Sichuan (SC) and Hunan (HN) accents.
More details are shown in Table \ref{tab:train}.

\begin{table}[!ht]
\renewcommand\arraystretch{1.0}
\caption{Details of both non-accented and accented English and Mandarin datasets. }
\centering
\scalebox{1.0}{
\begin{tabular}{lcc|lcc}
      \toprule
      \multicolumn{3}{c|}{English (\#hrs)}     &\multicolumn{3}{c}{Mandarin (\#hrs)}  \\
      \midrule
                  & Train        & Test         &
                  & Train          & Test           \\ \hline
       LibriSpeech  & 360    & 5.4      & Aishell   & 164      &  10   \\ 
       Indian       & 20     & 3.8      & Guangdong & 20       &  2.0    \\ \hline
       Canada       & -      & 2.2      & Hunan     & -        &  2.0     \\
       England      & -      & 1.9      & Sichuan   & -        &  2.0    \\
       \bottomrule
\end{tabular}}
\vspace{-0.4cm}
\label{tab:train}
\end{table}

\subsection{Experimental Setups}
\label{sec:setups}

We use 80-dimensional log Mel-filter bank (FBANK) plus
one-dimensional pitch as input acoustic feature. They
are computed using 25ms windows with a 10ms hop.
The utterance-level cepstral mean and variance normalization (CMVN)
computed using the training set is applied on the FBANK for
feature normalization. All our experiments are implemented with
the ESPnet ~\cite{Watanabe2018ESPnetES} end-to-end speech processing toolkit.
No data augmentation and no extra language model are applied.

For the acoustic encoder of Aformer, the input features are first
sub-sampled by the convolution subsampling module
which contains two 2-D convolutional layers with stride 2.
The general encoder of Aformer contains 12 conformer encoder layers
with 2048-dimension feed-forward and 256-dimension attention
with 4 self-attention heads. Two structures of accent encoder are
investigated, one is with 4 transformer encoder layers, the other
is with 2 layers 256-dimensional LSTM.
All models are trained with the Adam optimizer~\cite{Kingma2014adam}.
The warmup learning schedule~\cite{Gotmare2018ACL} is used for our
first 25K training iterations, and both label smoothing~\cite{Szegedy2016RethinkingTI}
weight and dropout is set to 0.1 for model regularization.

In English tasks, 3000 byte-pair-encoder (BPE)~\cite{Sennrich2015NeuralMT} units
generated by SentencePiece~\cite{Kudo2018SentencePieceAS} are taken
as the decoder outputs. In Mandarin tasks, we use 4231 Chinese characters
as the Aformer modeling units. The CTC weight is set to 0.3 during
both model training and inference. The word error rates (WER) and
character error rate (CER) are used for evaluating the ASR performance
for English and Mandarin tasks, respectively.

\subsection{Results and Discussions}
\label{sec:result}

\subsubsection{Results on English Accented-ASR}
\label{sec:english result}

\begin{table}[th]
\renewcommand\arraystretch{1.0}
\caption{WER(\%) for low-resource English accented ASR task. }
\centering
\scalebox{0.8}{
\begin{tabular}{l|l|l|l|ccc}
\toprule
   \multirow{2}*{\textbf{ID}}  &\multirow{2}*{\textbf{Train}}   &\multirow{2}*{\textbf{System}} &\multirow{2}*{\textbf{Fusion}}  &\multicolumn{3}{c}{\textbf{Test Set}}     \\
   &          &      &     &\textbf{IN}     &\textbf{EN}   &\textbf{CA}   \\
   \midrule
   \textbf{E1}  & LibriSpeech  & Conformer  & -   & 78.1    & 36.4   & 20.7    \\
   \textbf{E2}  & Indian    &  \quad  +Finetune    & -   & \textbf{36.7} & \textbf{40.6} & \textbf{23.2}  \\
   \midrule
   \textbf{E3}  &   &\multirow{4}*{Aformer}   & Add-LSTM   & 34.0    & 40.2   & 22.7  \\
   \cline{4-7}
   \textbf{E4}  & LibriSpeech    &    & Add                & 33.1    & 32.5   & 17.6   \\
   \textbf{E5}  &  +Indian       &    & Concat             & 33.2    & 32.5   & 17.9   \\
   \textbf{E6}  &     &               & Cross-attention    &\textbf{32.8}   &\textbf{32.2} &\textbf{17.5}   \\
   \bottomrule
\end{tabular}}
\vspace{-0.4cm}
\label{tab:English}
\end{table}

Table \ref{tab:English} presents the performance of low-resource
English accented ASR systems. E1 is the Conformer system trained
only on 360 hrs LibriSpeech data, and E2 is finetuned from E1
using 20 hrs Indian accent speech. E3 to E6 are all Aformer systems
that trained using the combined LibriSpeech and Indian training data
with the proposed multi-pass training. However, in E3, the accent
encoder is 2-layers LSTM, while in E4 to E6, their accent encoder
are the 4 transformer layers but with different encoder output fusion
methods.

Comparing the results of E1 system, it's clear that the Conformer performance
is significantly degraded when it meets accented test speech, especially
when the heavy accent deviates far from the training data.
E.g. on Indian test set, the WER reaches to 78.1\%, while on test sets with
England and Canada accents, the WER numbers are greatly
reduced to 36.4\% and 20.7\%. This is because compared with IN accent,
the EN and CA are more close to the speaking style or
acoustics of the data in LibriSpeech. When comparing E1 and E2,
we see 53\% WER reduction on IN test set, even there are slight
performance degradation on the out-of-domain EN and CA test sets.
It indicates that the traditional finetuning is very effective to
obtain good results on low-resource accented ASR task. Thus, we take
E2 as our baseline.

E3 and E4 are used to compare different accent encoder structures.
We see that, E4 is much better than E3, this may due to
the acoustic modeling ability of transformer is much stronger than
LSTM, and 4-layers transformer has more parameters than the 2-layers
LSTM. Difference between E4 to E6 are only the three information fusion
methods to combine the embeddings of general and accent encoders.
It's clear that, there is no big performance gap between
using linear addition (Add) and concatenation (Concat),
and the proposed cross-attention fusion achieves the best results.
In conclusion, the proposed Aformer with all three fusion methods
can obtain much better results than baseline system of E2,
both on the in-domain and out-of-domain accented test sets.
And compared with E2, the best system E6 achieves 10.6\%,
20.6\% and 24.5\% relative WER reduction on IN, EN and CA accented test set,
respectively. It means that, our proposed Aformer is more effective
and has stronger generalization ability to out-of-domain accented
speech than the conventional finetuning.

\subsubsection{Results on Mandarin Accented-ASR}
\label{sec:mandarin result}

\begin{table}[!th]
\renewcommand\arraystretch{1.0}
\caption{CER(\%) on Mandarin accented ASR task. }
\centering
\scalebox{0.8}{
\begin{tabular}{l|l|l|l|ccc}
\toprule
   \multirow{2}*{\textbf{ID}}  &\multirow{2}*{\textbf{Train}}   &\multirow{2}*{\textbf{System}} &\multirow{2}*{\textbf{Fusion}}  &\multicolumn{3}{c}{\textbf{Test Set}}     \\
   &          &      &     &\textbf{GD}     &\textbf{HN}   &\textbf{SC}   \\
   \midrule
   \textbf{M1}  & Aishell  & Conformer  & -   & 53.0    & 54.8   & 51.6    \\
   \textbf{M2}  & Guangdong    &  \quad  +Finetune    & -   & \textbf{29.4} & \textbf{38.7} & \textbf{40.1}  \\
   \midrule
   \textbf{M3}  &   &\multirow{4}*{Aformer}   & Add-LSTM   & 26.6    & 35.1   & 37.5  \\
   \cline{4-7}
   \textbf{M4}  & Aishell    &    & Add                & 25.7    & 34.7   & 35.6   \\
   \textbf{M5}  &  +Guangdong       &    & Concat             & 25.3    & 34.7   & 36.2   \\
   \textbf{M6}  &     &               & Cross-attention    &\textbf{24.8}   &\textbf{34.0} &\textbf{36.0}   \\
   \bottomrule
\end{tabular}}
\vspace{-0.25cm}
\label{tab:Mandarin}
\end{table}

Table~\ref{tab:Mandarin} shows the performance on the low-resource
Mandarin accented ASR task. Similar as the system E1 to E6 in Table \ref{tab:English},
M1 and M2 are taken as the baseline systems, M3 to M6 are the proposed Aformer
with different accent encoder and information fusion methods.
Different from the observation in E1 results, the CERs of M1 on the GD,
HN and SC are almost at the same level, it tells us that, all these three
accents deviates far from the non-accented Aishell training data.
When M1 is finetuned by 20 hrs Guangdong accent data,
the performance on all three accented test sets are greatly
improved, even the improvement gain on in-domain GD is much
larger than the ones on other two out-of-domain test sets.

The findings in M3 to M6 are consistent with the ones that observed
in E3 to E6 from Table \ref{tab:English}, such as, the best results
are also achieved from the Aformer (M6) using transformer accent encoder
and cross-attention method for information fusion.
Compared with M2, system M6 obtains a relative CER reduction of 15.6\%,
12.1\% and 10.2\% on the GD, HN and SC accented test sets, respectively.

\subsubsection{Ablation of Multi-pass Training}
\label{sec:ablation}

\begin{figure}[h]
  \centering
  \includegraphics[width=8cm]{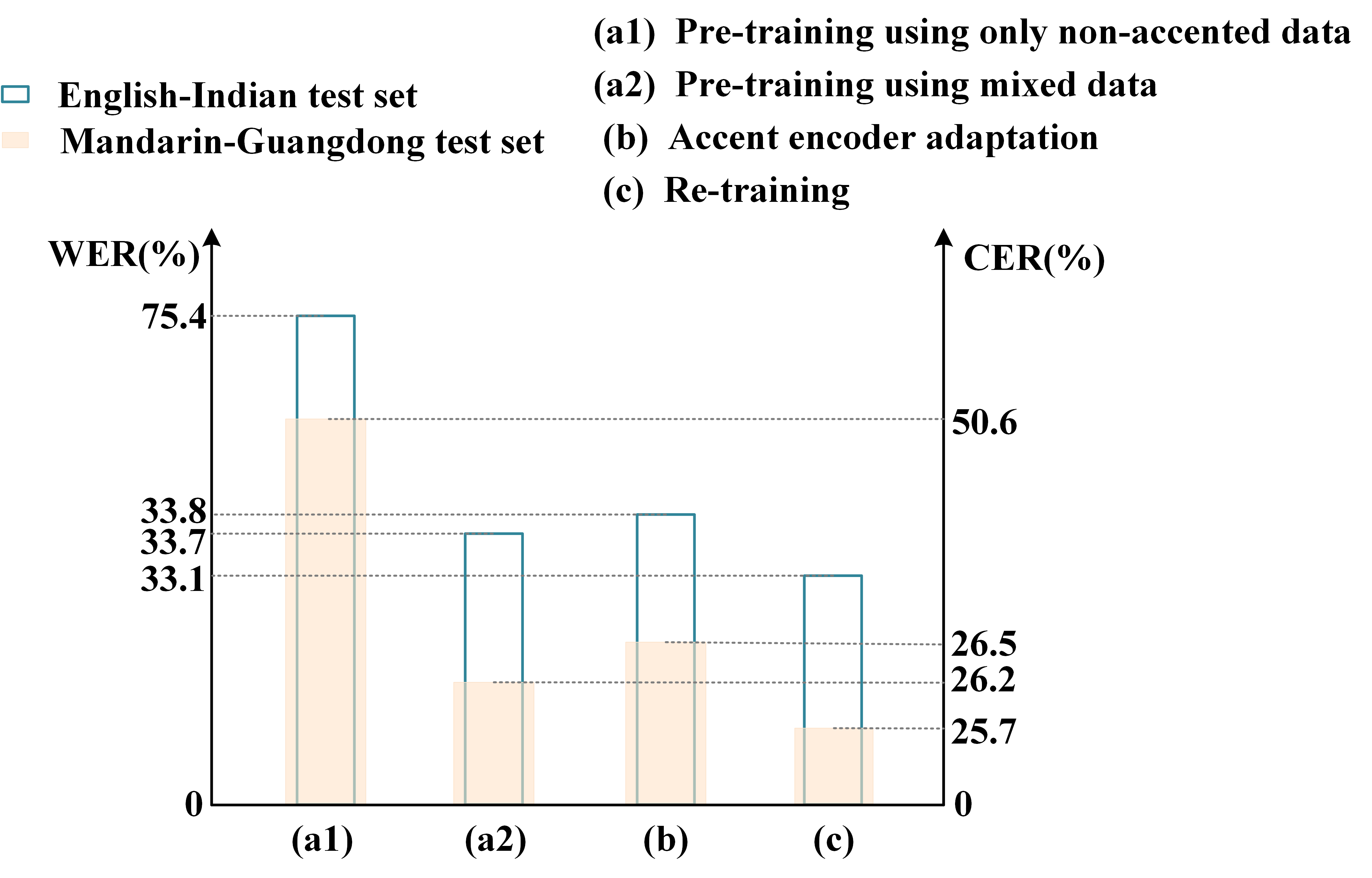}\\
  \caption{Ablation study of multi-pass training method.}
  \label{fig:Ablation}
\end{figure}

Fig.~\ref{fig:Ablation} shows our ablation experimental results 
to verify the effectiveness of the proposed multi-pass training. 
All these experiments are performed on E4 and M4. In 
Fig.~\ref{fig:Ablation}, four bars means using 
the three different training stages in multi-pass training that 
described in Section \ref{sec:Multi-pass Training}. It is worth mentioning that 
(a1) is the "Pre-training" described in Section 3.2 with only non-accented data, and (a2) is the pre-training with mixing all the accented and non-accented data together 
to train the Aformer structure. (b) and (c) are exactly the 
same as described in Section \ref{sec:Multi-pass Training}. The white four 
bars are WER\% on the Indian test set of E4, while the pink four ones are CER\% on the Guangdong test set of M4.    
It's clear that, the re-trained Aformer achieves 
the best results on both English and Mandarin accented ASR tasks.
It means that, the multi-pass training is effective than only 
using pooling data to train the Aformer, and the Aformer that 
finetuned using the accented training data.

\vspace{-0.5cm}

\section{Conclusion}
\label{sec:conclution}

In this study, we explore the approach of leveraging large amount of non-accented 
training data to enhance the low-resource accented end-to-end ASR system. 
Based on the standard Conformer ASR architecture, we propose an Aformer 
to capture both the general acoustic context and accent-dependent 
acoustic information. Moreover, a multi-pass training and different 
cross-information fusion methods are also investigated to further 
improve the Aformer. Results on both the low-resource accented English and Mandarin 
ASR tasks show that, the proposed methods outperform 
the finetuned Conformer significantly, either on the in-domain or the 
out-of-domain accented speech test sets.

\bibliographystyle{IEEEtran}
\bibliography{mybib.bib}

\end{document}